\documentclass[]{article}%%%%where rstrans is the template name
\usepackage{graphicx}
\usepackage{amsmath,amssymb}

\newcommand\md{\mathrm{d}}

\bibstyle{unsrt}
\title{Fury: an experimental dynamo with anisotropic electrical conductivity}

\author{T. Alboussière$^{1}$, F. Plunian$^{2}$ and M. Moulin$^{3}$}

\date{$^{1}$Universit\'e Lyon 1, ENS de Lyon, CNRS,
 Laboratoire~de~G\'eologie~de~Lyon,\\ 69622~Villeurbanne,~France\\
$^{2}$Universit\'e Grenoble Alpes, Universit\'e Savoie Mont Blanc, CNRS, IRD, IFSTTAR, ISTerre,\\  38000 Grenoble, France\\
$^{3}$ENS de Lyon, CNRS, Laboratoire de Physique,\\ 69364 Lyon, France}

\begin{document}

\maketitle

%%%% Abstract text to be placed here %%%%%%%%%%%%
\begin{abstract}
	We report measurements of dynamo action in a new experimental setup, named Fury, based on the use of an anisotropic electrical conductivity. It consists in a copper rotor rotating inside a copper stator, electrically connected with a thin layer of liquid metal, galinstan. Grooves have been cut in the copper so that, everywhere, electrical conductivity can be considered to be that of copper along two directions while it is zero along the third one. The configuration is efficient and dynamo action can be powered by hand. We have also used a motor with better control, enabling us to drive the rotor at specified velocity or torque functions of time. The structure of the axisymmetric magnetic field produced is found to be close to the numerical modelling using FreeFem++. The experimental dynamo behaves very nearly as expected for a kinematic dynamo, so that the threshold dynamo velocity cannot be exceeded, or only briefly. More mechanical power in the rotor rotation leads to an increase in the magnetic field intensity, the magnetic energy being proportional to the extra mechanical power beyond threshold. In the transient following a step increase of torque, magnetic and angular velocity oscillations have been observed and explained.
\end{abstract}
%%%%%%%%%%%%%%%%%%%%%%%%%%%

%%%%%%%%%% Insert the texts which can accomdate on firstpage in the tag "fmtext" %%%%%

\section{\label{intro} Previous dynamos}
%%%% Insert A head here

The notion "dynamo" can refer to different machines and was historically used to designate a device transforming mechanical energy into continuous electric current. 
Faraday's dynamo was the first \cite{faraday}: a disk rotating in the magnetic field generated by a permanent magnet leads to an electric potential difference between its centre and rim and to an electric current when they are connected. This fundamental discovery - induction - was then used in more powerful machines to produce electricity from mechanical energy (the names of Pixii in France, Pacinotti in Italia, Woolrich in England are associated with such machines). However, they needed many heavy magnets to produce a modest electric power. This was at the origin of the concept of "sef-exciting dynamos". The idea was to reinforce (or excite) the magnetic field of the permananet magnets by diverting a fraction of the electric current produced by the machine and wind it around the magnets, thus becoming electro-magnets. \`Anyos Jedlik, around 1856 in Gy\"or in the kingdom of Hungary (at the time), seems to have been the first inventor of such a machine, but his work remained unknown for a long time. He was followed by Hjorth in Danemark, Siemens in Germany, Gramme in France, Heatstone and Wilde \cite{wilde1867} in England. Note, however, that the self-exciting dynamos needed some residual magnetization in their iron parts when they were started: initially the motion of the rotor in the residual magnetic field produces a small electric current that is used to feed the electromagnets, so that the same rotation rate produces more current up to the point when part of it can be used to power an external load. This is so true that a well-known cause of failure is the lack of residual magnetism sometimes due to a long period of inactivity. In such a case, the dynamo will not produce electricity even at high rotor speed and the solution known as "flashing the field" consists in having an external generator, or a battery, discharge electricity in the windings of the electromagnets to restore residual magnetism. Even now, this solution can be found on the web when a car alternator does not work. Another hint of the importance of residual magnetization comes from visits of electricity museums: even after many decades of inactivity, all these self-exciting dynamos have a magnetic field of the order of 1000~Gauss, as can be measured by anyone with a smartphone and the excellent application PhyPhox \cite{phyphox} giving access to all sensors (acceleration, sound, pressure, magnetic field) present in these mobile devices.

Here, by "dynamo", we mean a device or a natural object producing a magnetic field from the motion of electrically conducting materials, in the absence of magnetized materials (or materials with magnetic properties in the strictest acceptance). Many planets and all stars have a self-generated magnetic field, whereas their internal temperature is above the Curie point, precluding any magnetic effect. In this context, the corresponding concept of self-exciting dynamo of Bullard \cite{bullard} differs definitely from the historical definition mentioned above. Self-excitation now refers primarily to the capacity of the dynamo to operate in the absence of any initial magnetic field. We expect -- and this is true for Bullard dynamo -- that an instability threshold is reached when the rotation rate is large enough, above which the magnetic field and electric current will jointly grow simultaneously exponentially (initially). This corresponds to a supercritical bifurcation, which is the case in most self-exciting dynamo models in the modern geophysical and astrophysical sense. In contrast, we formulate the hypothesis that the historical self-exciting dynamo machines of the 19th century were probably dynamos in a subcritical regime, with an instability threshold above the usual range of nominal rotation rates. Alternatively, their different behaviour could be due to the non-linearities: a residual (remanent) magnetic field is a positive non-linearity when it is present, however contact resistance at sliprings constitutes a very negative non-linearity for dynamo action. Magnetic saturation is also a non-linear effect but should not impact the starting phase of dynamos.

Concerning modern experimental realizations of self-exciting dynamos, with solid or liquid electrical conductors, 
 there have been four self-generated experimental dynamos published in the literature. One is a `solid' dynamo \cite{Lowes1963,Lowes1968}, the other three have used liquid sodium \cite{Gailitis2001,Stieglitz2001,Monchaux2007}. Soon after the first theoretical dynamo model \cite{herzenberg}, Lowes and Wilkinson \cite{Lowes1963} made an experimental version of it, consisting in two cylindrical electrically conducting rotors, rotating in cylindrical bores in a solid sphere (or a cube in the second version) and in electrical contact through a film of liquid metal (mercury). Both rotors and the sphere have been made of soft iron to increase the magnetic permeability by a factor $250$ ($150$ in the second version), in order to lower the dynamo threshold in terms of angular velocities. Dynamo action was reached with approximatively $200$~W of mechanical energy. It took time and effort before fluid dynamos were realized. In 2000, the experiments in Riga \cite{Gailitis2001} and Karlsruhe \cite{Stieglitz2001} showed that dynamo action was possible with liquid sodium. In Riga, a flow inspired from the Ponomarenko model \cite{Ponomarenko1973} was set up: a turbine propelled the fluid with strong rotation in a pipe, the fluid was then channelled back around the pipe while its rotation was stopped using baffles \cite{Gailitis2001}. Dynamo action was obtained with $100$~kW and $1.5$ cubic meter of sodium. In Karlsruhe, an arrangement of helical pipes produced an array of helical flows of the same helicity sign. This is an approximation for the G.O. Robert's model \cite{Roberts1972}. With $1.6$~m$^3$ of sodium and about 500~kW, a stationary magnetic field was produced \cite{Stieglitz2001}. Finally, in Cadarache, the VKS experiment produced a dynamo effect in a cylinder with two counter-rotating discs at each end \cite{Monchaux2007}. This arrangement can be seen as a realization of a Taylor-green vortex, although the flow becomes highly unsteady and turbulent for a moderate rotation rate of the discs. The discs had to be made of iron in order to reach the dynamo threshold. This setup required 160~liters of sodium and $150$~kW for dynamo action.

With the more restrictive definition excluding magnetic properties, only two of the four experiments are dynamos: Riga and Karlsruhe dynamos. Their mean flow is well-constrained and their dynamo threshold was predicted correctly from the kinematic dynamo analyses. Of course, above the threshold, the flow was altered by the magnetic field and a non linear dynamo regime was observed \cite{Gailitis2003, Tilgner2001}, involving a change in the structure of the flows.  On the contrary, the Cadarache dynamo is less constrained in terms of flow structure. The dynamo threshold was not predicted from the mean flow and a larger variety of non linear behaviours have been observed \cite{berhanu2007}. In all three cases, the mechanism of saturation above the threshold is not completely understood \cite{Gailitis2001,muller2004,muller2004b}. This is partly due to the fact that the setups have a limited amount of mechanical power and cannot reach velocities much above threshold, and partly due to the difficulty to detect very subtle changes in the velocity field.

\section{\label{Model} Fury: from copper and grooves}

\begin{figure}[!h]
        \begin{center}
                \includegraphics[height=6.4 cm, keepaspectratio, angle=0]{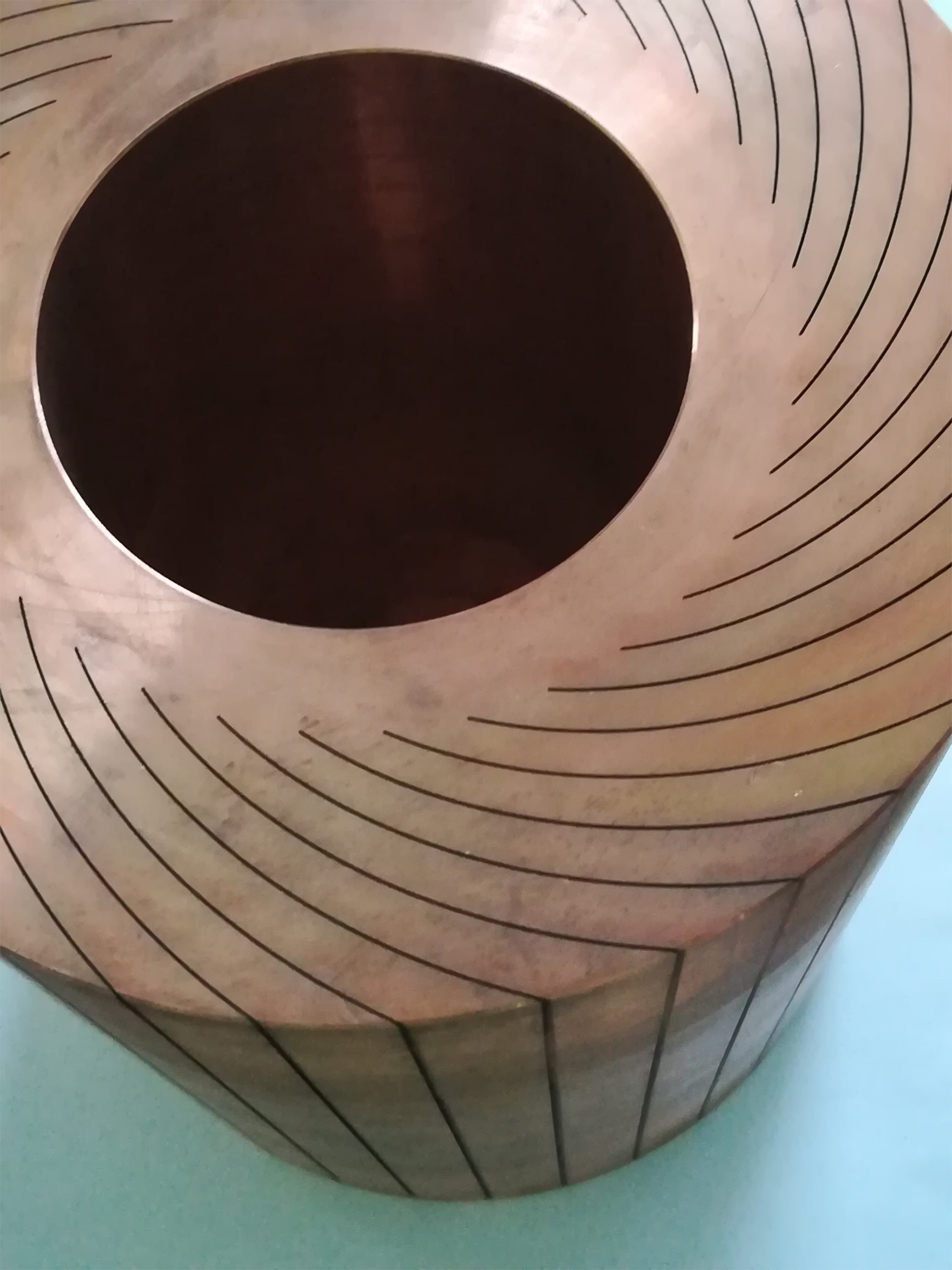}\hspace{4 mm}\includegraphics[height=6.4 cm, keepaspectratio, angle=0]{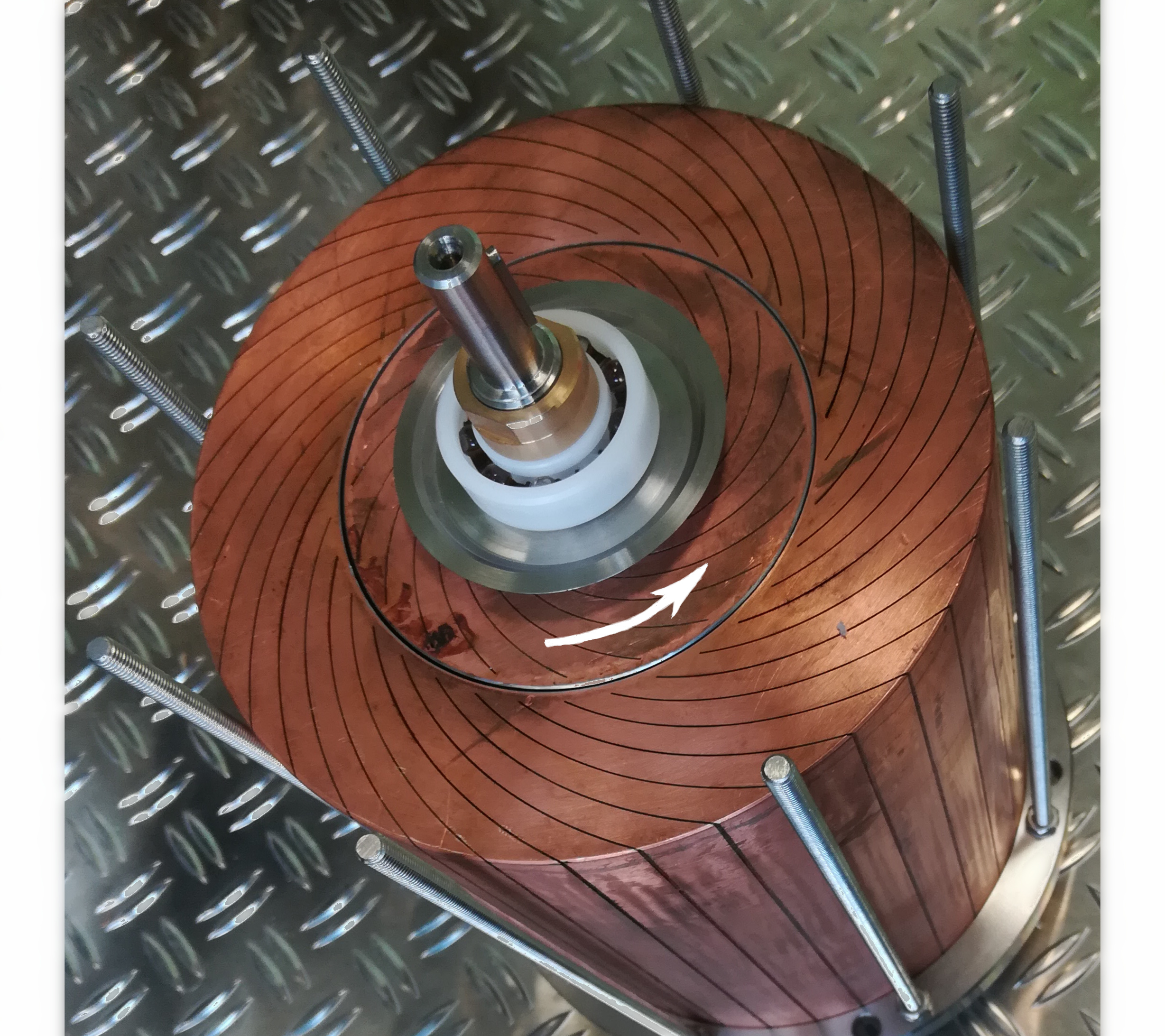}
                \caption{Grooves are cut in a copper rotor and stator.}
                \label{fig_rot_stat}
        \end{center}
\end{figure}

From our previous theoretical investigations on anisotropic electrical conductivity \cite{Alboussiere2020,Plunian2020}, we know that a dynamo can have a planar velocity field and/or produce an axisymmetric magnetic field. We have built an experimental setup which is a rotor inside a stator (Fig.~\ref{fig_rot_stat}). The final external diameter of the stator is 170~mm, its length is 209~mm and its internal diameter is 101~mm. The rotor that fits inside has the same length and an external diameter of 100~mm, hence the gap between rotor and stator is $0.5$~mm wide. It has an internal diameter of 30~mm so that a (non-magnetic) steel axis can be fitted in. The grade of copper for rotor and stator is CUA1 H12, with a guaranteed electrical conductivity $\sigma$ between $100$ and $102.5\ \%$ IACS, {\it i.e.} $\sigma = 5.8\ 10^7 \, \Omega ^{-1}$m$^{-1}$ to $\sigma = 5.95\ 10^7 \, \Omega ^{-1}$m$^{-1}$.

Grooves have been cut in the rotor and stator, extending across the whole length and forming arcs of logarithmic spirals in the perpendicular cross-section: the direction along the groove makes an angle $62\ ^{\circ}$ with the radial direction at all points. In the stator 35 equally-spaced spirals have been cut and 20 in the rotor. The grooves have been cut with the technique of electro-erosion, each groove has a width of $0.33$~mm. The grooves have then been filled with a polyimide film and epoxy resin for electric insulation and mechanical strength. As can be seen on the left-hand side of Fig.~\ref{fig_rot_stat}, the internal diameter of the stator was cut 20~mm smaller initially to ensure rigidity during the cutting of the grooves (similarly, the external diameter of the rotor was larger by 20~mm). After the epoxy resin hardened, the stator internal and rotor external diameters were cut to the final dimensions, leaving a radial distance of 2~mm to the end of the grooves. This ensures a good precision and small roughness of the surfaces involved in the differential rotation. This surfaces (external of the rotor and internal of the stator) were then silver plated (a few microns) for chemical protection of the copper and to ensure good electrical contact with galinstan.

\begin{figure}[!h]
        \begin{center}
                \includegraphics[height=8.0 cm, keepaspectratio, angle=0]{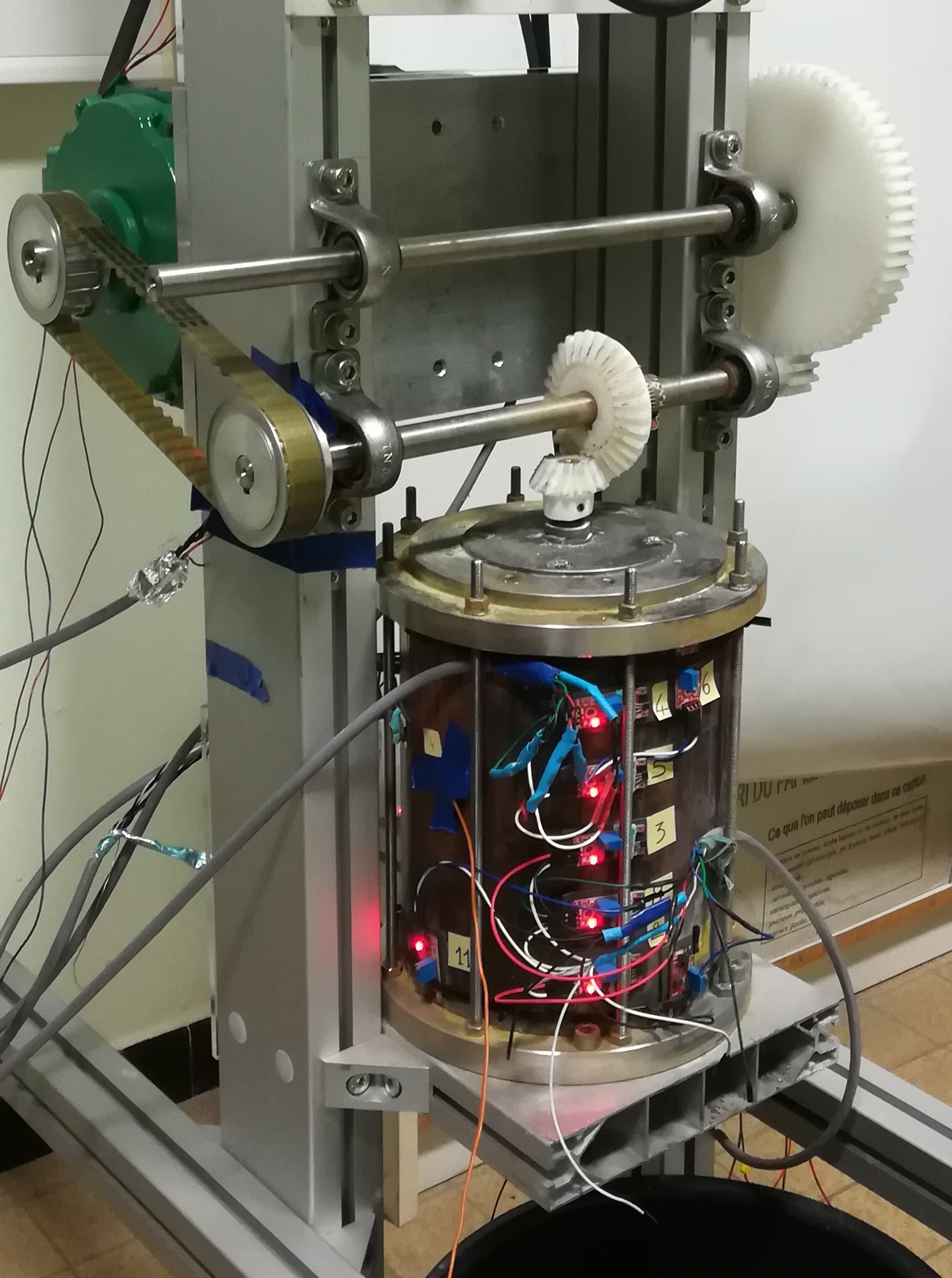}\hspace{4 mm}\includegraphics[height=8.0 cm, keepaspectratio, angle=0]{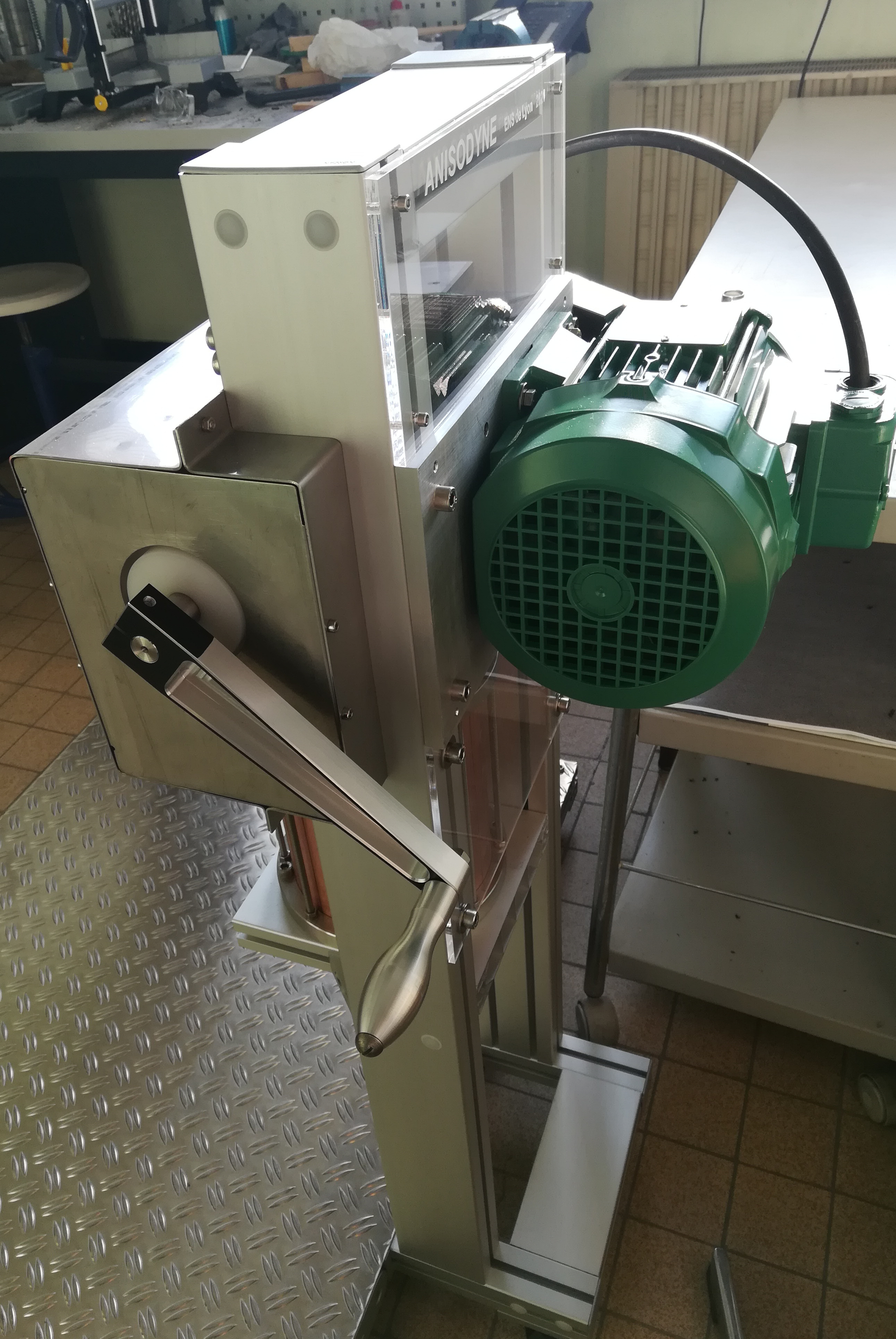}
                \caption{Front view of Fury (left) and side view (right).}
                \label{view}
        \end{center}
\end{figure}

The assembly of the setup is quite simple, with two flanges at the top and bottom tied with threaded rods, ensuring mechanical cohesion with ball bearings inserted to hold the axis of the rotor. The 0.5~mm gap between rotor and stator was filled with galinstan, an alloy of gallium, indium and tin, liquid at room temperature.
In Fig.~\ref{view} showing the final setup, the copper stator is still visible and the rotor axis is connected through gears and belt to a handle and to an electric motor. Manual or motor operations are possible. One turn of the handle (resp. electric motor) corresponds to 12 (resp. 2) turns of the rotor axis.

%The 0.5~mm gap between rotor and stator was filled with galinstan, an alloy of gallium, indium and tin, liquid at room temperature. At a typical rotation rate of 24~Hz, the Reynolds number in the gap is approximately equal to $10^4$. XXXXXX

Measurements are made of the magnetic field from Hall probes on the external surface of the stator: each Hall probes has a small red LED, see Fig.~\ref{view}. The electric variator of the motor provides measurement of angular velocity and torque.

\section{\label{expfury} Fury: dynamo results}

The dual operation, manual or motor-powered, has two distinct objectives. Powering the experiment by hand is interesting to have a direct physical feeling of dynamo action: one can feel the mechanical resistance due to Lorentz forces when the magnetic field is generated. However, the conditions of operation are not well controlled: the velocity is not constant nor the torque and is operator-dependent. Using a motor brings the possibility to follow almost any prescribed function of velocity or torque, within the limit of its maximal power.

\subsection{\label{first}The first successful human-powered dynamo run}

\begin{figure}[!h]
        \begin{center}
                \includegraphics[height=6.5 cm, keepaspectratio, angle=0]{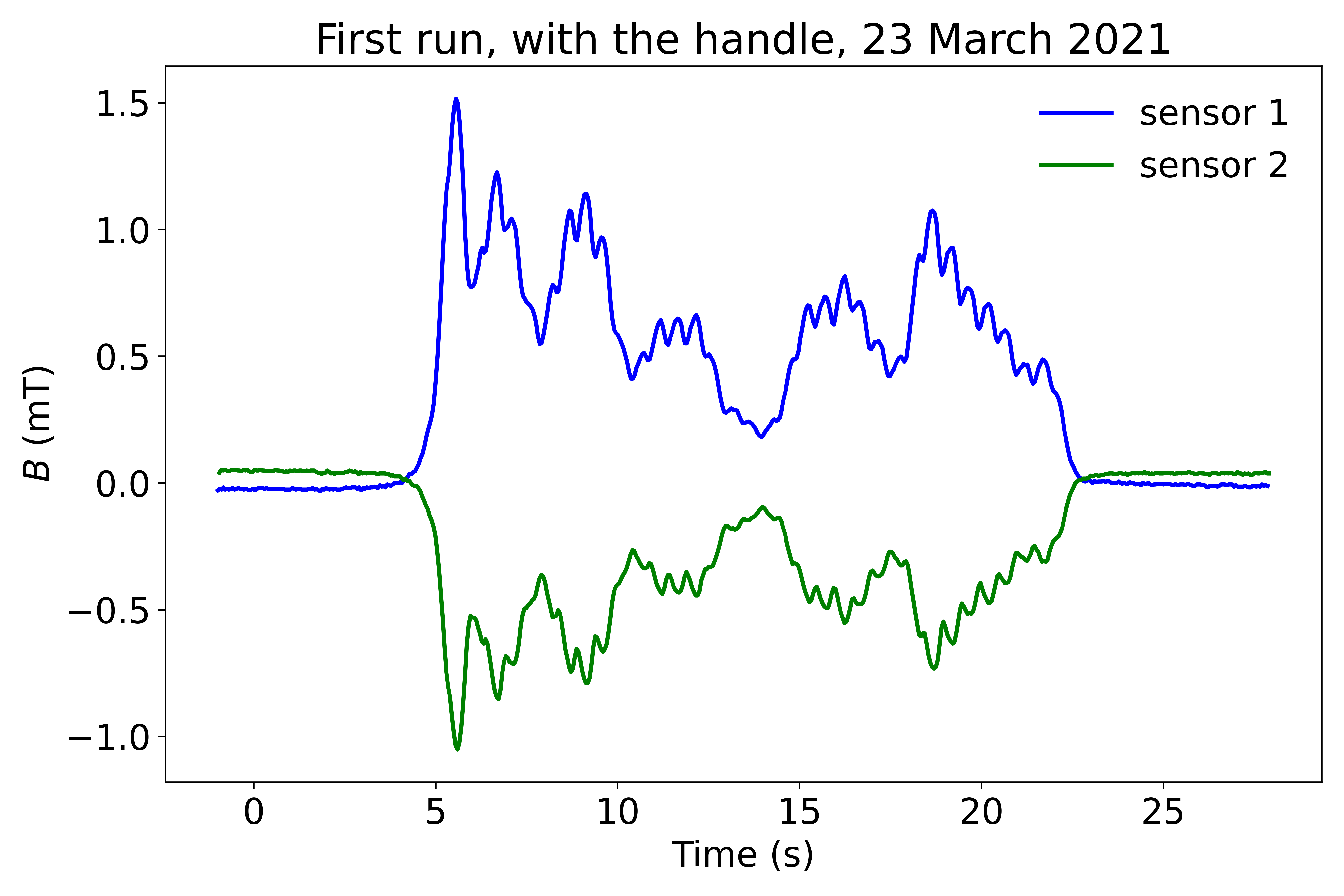}
                \caption{First dynamo action with Fury, human-powered with the handle.}
                \label{firstrun}
        \end{center}
\end{figure}

The very first run of dynamo action with Fury took place on the 23rd of March 2021. In Fig.~\ref{firstrun}, one can see the signal of the two Hall probes (at the time) stuck on the outside of the stator. Those two signals are proportional to each other, although the two probes were at different places and had different orientations. This hints at the fact that the self-generated magnetic field is made of a single mode. Only its magnitude can change with time. Let us analyze the time evolution of this run. The operator ran the handle until it reached about 2 Hz, at a time between $t=2$~s and $t=3$~s.  The magnetic field started to depart from its initial value (the ambient magnetic field). At the time nearly $t=6$~s it peaked to a maximum. In the same time, the operator had experienced an extra resistance from the Lorentz forces, which led to a decrease of the rotation rate and irregular decay of the magnitude of the magnetic field. At $t=13$~s, the operator recovered a little and put more effort for a while, then stopped at $t=22$~s or so.

Another information from the signals in Fig.~\ref{firstrun} is the presence of a 2~Hz oscillation. This is precisely the frequency of the handle rotation. The interpretation is that the input of energy is not constant over one rotation of the handle, which seems quite plausible. Note that the ratio between handle and rotor rotation is 12 (a factor 6 in the gears between the horizontal axes, see Fig.~\ref{view}, and 2 in the conical gears on top of the rotor). This implies that the threshold for dynamo action corresponds to an angular velocity around $24$~Hz for the rotor.

As will be seen later, reaching 2~Hz with the handle requires an input of $230$~W of mechanical work to overcome the resistance of the galinstan film (main contribution), ball bearings and gears. This is quite an effort with a handle, but all operators over a representative sample of ten colleagues (two women, eight men) could do it. The good news is that the effort does not need to be sustained for a long time. Within 5~s, it is usually possible to start the dynamo.

\subsection{\label{motorresults}More results with a motor}

Most results have been obtained with an electric motor of maximal power 750~W (about one horse power). This is just a little bit more than what the average human can do, but it can sustain it for longer and can adjust the effort very precisely.

\subsubsection{\label{eigenvector}Marginal eigenvector}

\begin{figure}
        \begin{center}
                \includegraphics[height=9.7 cm, keepaspectratio, angle=0]{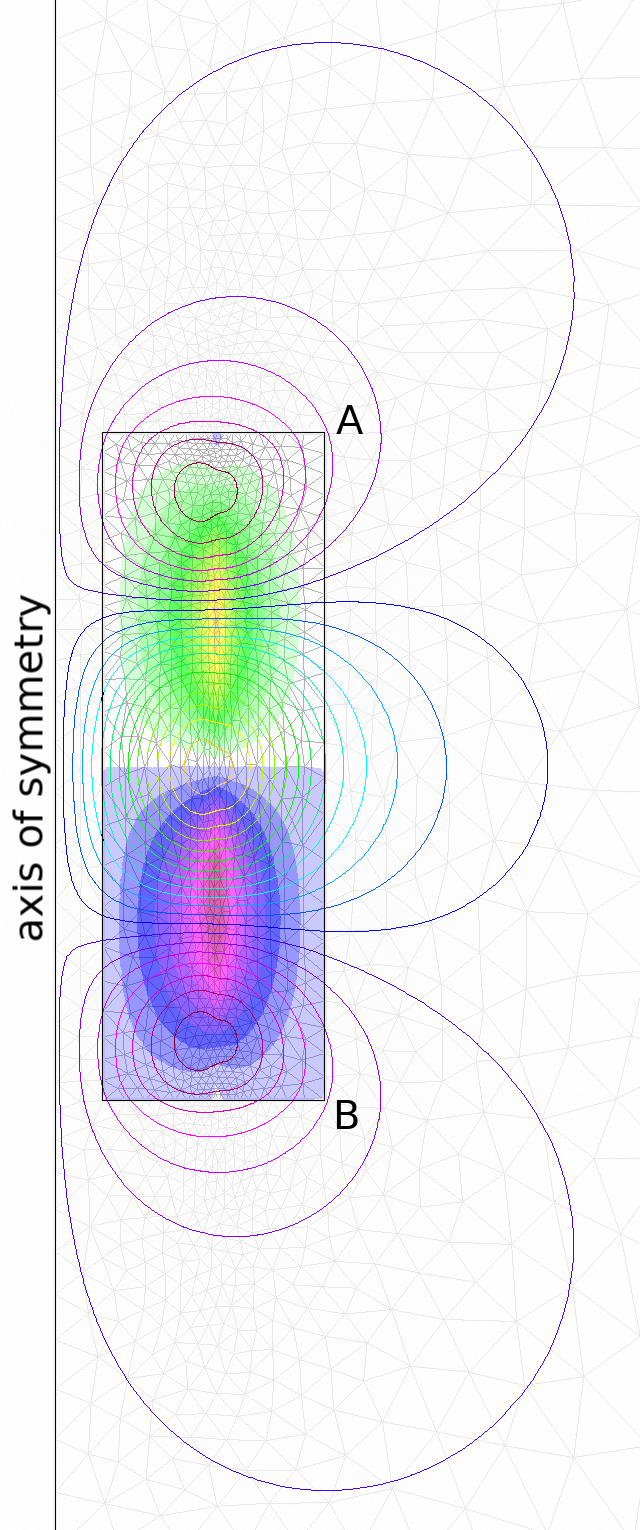}\hspace{4 mm}\includegraphics[height=9.7 cm, keepaspectratio, angle=0]{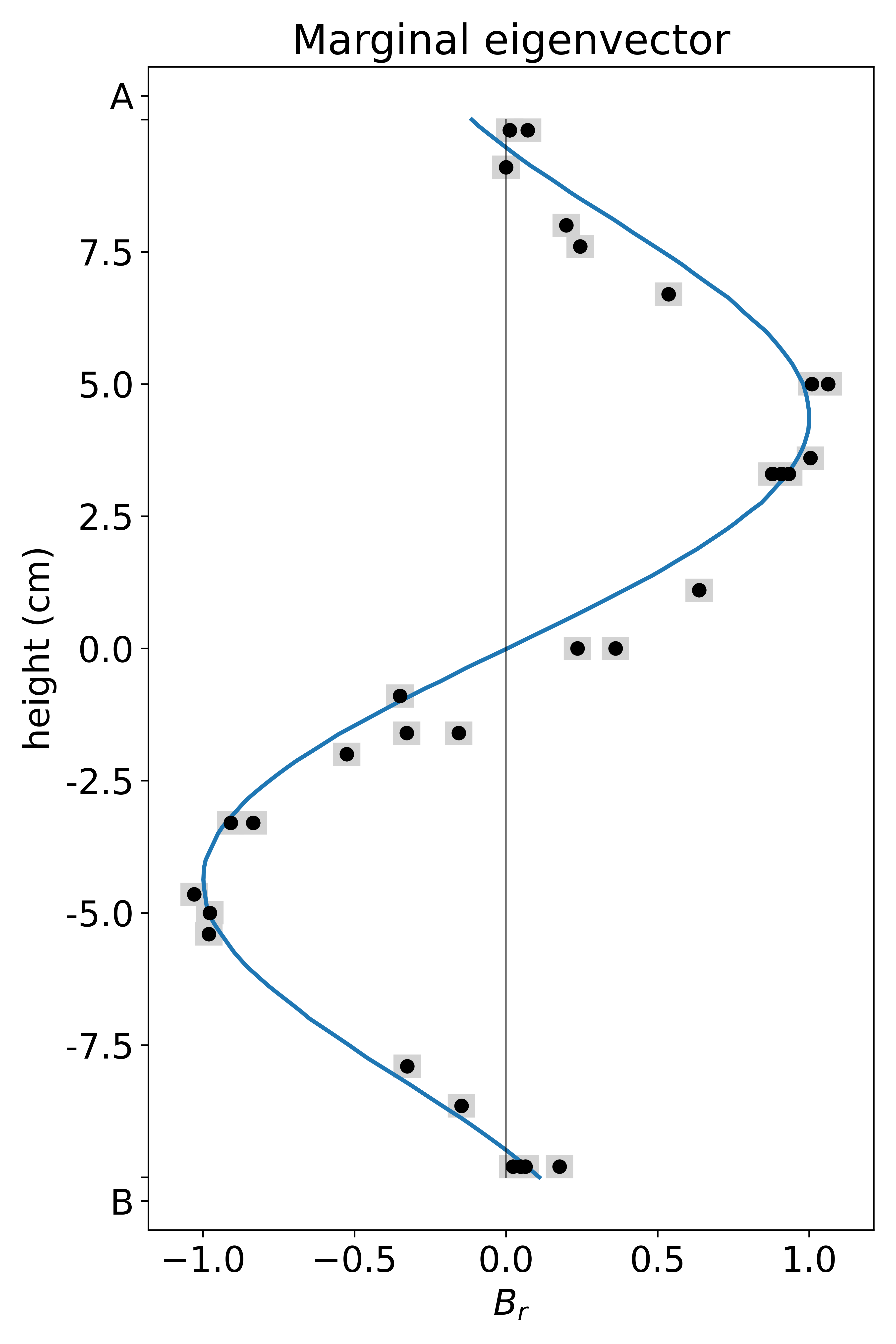}
                \caption{Eigenmode at marginal stability from finite element modelling (FreeFem++) on the left, profile of $B_r$ (FreeFem++ and measured on Fury) on the right.}
                \label{mode}
        \end{center}
\end{figure}

Using FreeFem++ \cite{freefem}, we have built a numerical model of Fury. This model is assumed to be axisymmetrical as the magnetic field observed in the experiment is itself axisymmetric. Electrical conductivity is taken to be anisotropic in the coper parts. This means that we have not attempted to model each copper blade, but have used a mesoscale anisotropic modelling instead. The finite-element model takes into account the thin galinstan gap with its isotropic electrical conductivity about $16.8$ times smaller than that of copper. The induction equation is considered, with a solid rotation of the rotor, and a linear profile in the galinstan gap to reach a zero velocity in the stator. Outside the conducting domain, the magnetic field obeys a harmonic equation.   

We have computed the marginal eigenvalue and eigenvector at the lowest possible rotation rate. This (axisymmetric) eigenmode is shown on Fig.~\ref{mode}: the mesh used for the finite element calculations is shown in the background, the toroidal isovalues are shown with colour level in the conducting domain (rotor-galinstan-stator) and some field lines of the poloidal component are drawn. On the right-hand side of Fig.~\ref{mode}, the cylindrical radial component of the magnetic field is plotted along the profile {\tt{A}-\tt{B}} on the outside surface of the stator (blue curve). The black dots are measurements of the same profile on Fury. The grey region around each point corresponds to the uncertainty of position ($\pm\ 2$~mm and magnetic field $\pm 5$~\% of the maximum value, mainly due to orientation errors of Hall sensors). These points correspond to different runs and to different azimuthal angles. Considering the accuracy of the measurements, there is no sign of departure from axisymmetry. The agreement between the calculated profile and the measurements is quite good: it seems that the experimental values are slightly shifted to the bottom compared to the calculated odd profile.

Let us now consider the value of the magnetic Reynolds number corresponding to marginal stability.
Let us define the magnetic Reynolds number as $Rm = \mu \sigma U R$, with $\sigma$ the electrical conductivity of copper, $\mu = 4 \pi \, 10^{-7}$~H~m$^{-1}$ the magnetic permeability of copper (or of vacuum), $R=0.05$~m the radius of the rotor and $U=\omega R$ the maximal tangential velocity of the rotor ($\omega$ is the angular velocity). A rotation rate of the rotor of $24$~Hz corresponds to an angular velocity of $150$~rad~s$^{-1}$. However, with the motor and more precise velocity measurements, we could get dynamo action for $\omega \simeq 147$~rad~s$^{-1}$ (red curve in Fig.~\ref{couplevelocity}). This leads to a critical magnetic Reynolds number of $Rm_c \simeq 26.8$. The dynamo threshold obtained with FreeFem++ is $Rm_c \simeq 23.1$. The difference can be explained firstly by the presence of a finite number of discrete grooves in the experiment, compared to a uniform anisotropy in FreeFem++ calculations. Secondly, the grooves correspond to a removal of about 5~\% of the total mass of copper. This would already explain an increase by 5~\% of the threshold. There can also be a contact electrical resistance between the copper and the film of galinstan. For completeness, the analytical threshold is $Rm_c \simeq 14.61$ with an infinitely long configuration and a stator of infinite radius (and in the absence of a layer liquid galinstan of electrical conductivity $16.8$ times smaller than that of copper) \cite{Plunian2020}.

In accordance with theoretical investigations \cite{Alboussiere2020,Plunian2020}, when the imposed rotation has the opposite direction, dynamo action never happens: all eigenvalues from FreeFem++ have a negative real part. Similarly, no dynamo action can be seen in Fury with the negative rotation. 

\subsubsection{\label{power}Torque, power and magnetic energy}

\begin{figure}
        \begin{center}
                \includegraphics[height=7.5 cm, keepaspectratio, angle=0]{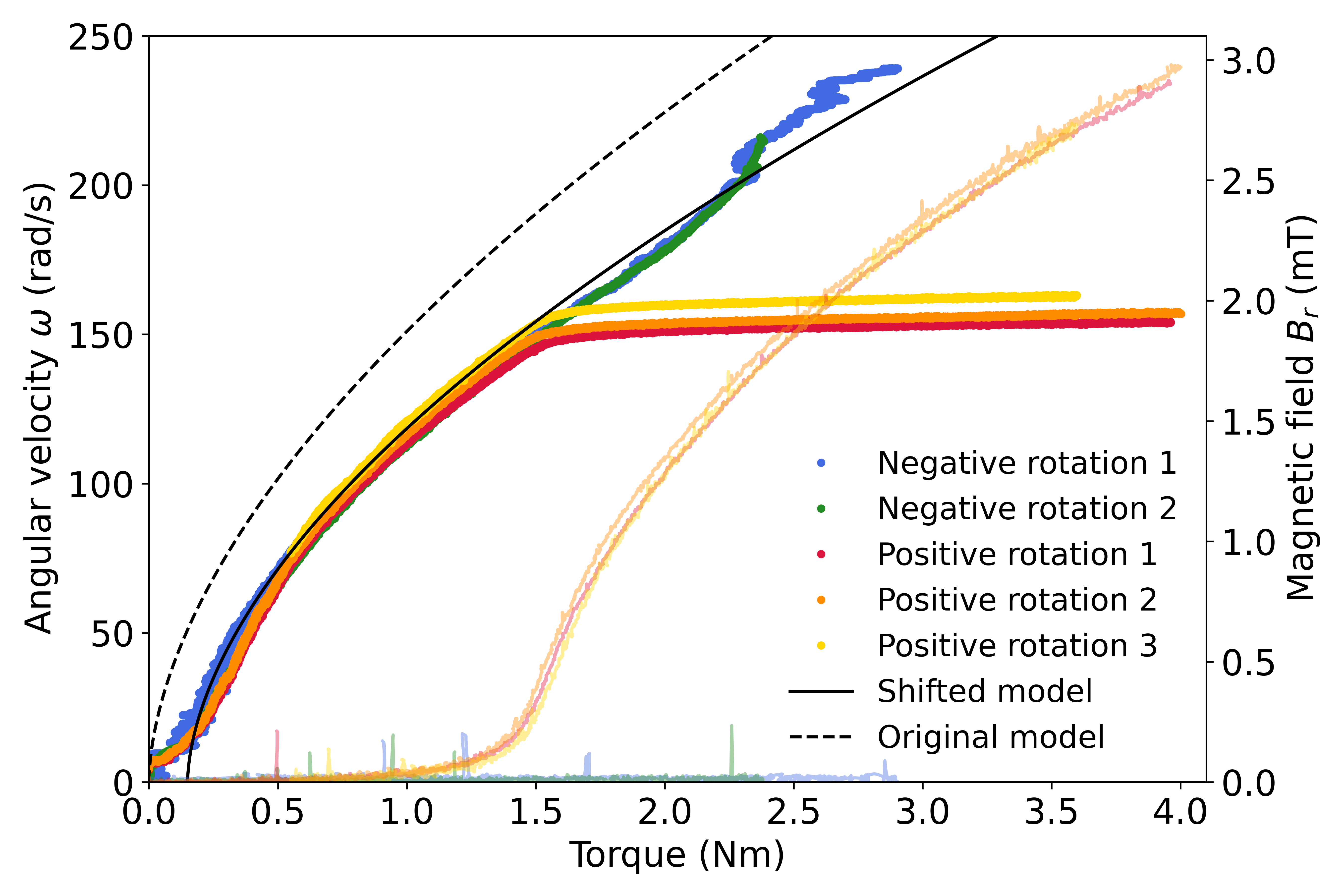}
		\caption{Angular velocity (bright colours) and magnetic field (faded colours) versus torque, for two runs in the negative direction and three runs in the positive (dynamo) direction. }
                \label{couplevelocity}
        \end{center}
\end{figure}

Increasing linearly the imposed torque from zero to its maximum value, we have been able to plot the resulting curves in the diagram shown in Fig.~\ref{couplevelocity}. Three runs in the ``positive'' direction of rotation are shown, and two in the reverse ``negative'' direction (see Fig.~\ref{fig_rot_stat} where the white arrow defines positive rotation). In agreement with the mathematical analysis \cite{Alboussiere2020,Plunian2020} dynamo action is possible only with the positive direction of rotation. At low rotation rate, below the dynamo threshold, all curves follow the same path. The relationship between the angular velocity and the torque is determined by hydrodynamical properties of the galinstan alloy with additional torque contributions from the ball bearings (outside on the visible axes on Fig.~\ref{view}, inside to support the axis of the rotor as shown in Fig.~\ref{fig_rot_stat}) and gears. The dashed black curve (original model) in Fig.~\ref{couplevelocity} corresponds to the turbulent model by Orlandi {\it et al.} \cite{orlandi}\footnote{A fit for the friction coefficient $c_f$ is proposed for the turbulent Couette flow in terms of the Reynolds number $Re$, based on various experimental results $c_f = 0.056 Re^{1/4}$. The Reynolds number is $Re = U h / \nu$, where $U$ is half the velocity difference across a layer of thickness $h$ of a fluid of kinematic viscosity $\nu$. The resulting shear stress is $\tau _w = \frac{1}{2} c_f \rho U^2$ by definition of the friction coefficient $c_f$.}. Note that at a typical rotation rate of 24~Hz, the Reynolds number of the galinstan in the gap is approximately $5000$ (plane Couette flow is generally turbulent above Reynolds number $10^3$). The continuous black curve (shifted model) is derived from the first one with a multiplication by $1.3$ and with the addition of a constant resistive torque of $0.15$~Nm, for a best fit of the data. No attempt was made to measure the fluctuations of the torque (for an imposed velocity) or of the velocity (for an imposed torque) that would be due to hydrodynamic turbulence in the gap of galinstan. We expect those fluctuations to produce a small average effect (over the whole gap) and to be impossible to measure from the variator of the electric motor. 

In the positive direction of rotation, above $150$~rad~s$^{-1}$, increasing further the torque does not result in an increase in angular velocity. The self-generated dynamo behaves pretty well like a kinematic dynamo: it is virtually impossible to exceed the threshold angular velocity ($150$~rad~s$^{-1}$ corresponds to $150/(2 \pi) \simeq 24$~Hz). Increasing the torque leads to an increase in the self-generated magnetic field (fade colored curves in Fig.~\ref{couplevelocity}).  On the contrary, in the negative direction of rotation, the same ``hydrodynamical'' curve is followed as for lower velocities and no magnetic field is generated.

In the dynamo regime (positive direction), a careful examination reveals that the angular velocity increases very slightly when the torque is increased above the threshold. However, most of this increase is due to the effect of temperature. As the system dissipates energy (by turbulence and viscosity at small scale, by Joule heating when the dynamo is running) the copper of the rotor and stator gets heated hence its electrical conductivity decreases. In order to maintain the kinematic dynamo at its critical magnetic Reynolds number, a decrease in electrical conductivity has to be compensated by a similar relative increase in angular velocity. In Fig.~\ref{couplevelocity}, the three dynamo runs, 1 to 3, were obtained while the temperature of the outside surface of the stator was 20.2, 22.4 and 25.6$^\circ$C respectively. With a linear thermal coefficient of $4.3\ 10^{-3}$~K$^{-1}$ for copper, the difference in temperature of 5.4~K leads to a relative variation of 2.3~\% in electrical conductivity. The relative difference in angular velocity between run 1 and run 3 is about 5~\%. We do not have a definite explanation for this mismatch: one possibility is that the temperature difference in the interior of the setup was larger than that measured on the external surface.

Anticipating on the next section \ref{stepssection}, we can see in Fig.~\ref{stepsenergy} that the magnetic energy grows linearly with the input mechanical power injected in the setup, when the dynamo threshold is exceeded. In that regime, any extra power input is dissipated by an extra Joule dissipation. This is coherent from the point of view of magnetohydrodynamics, as Joule dissipation scales usually proportionally to the square of the magnetic field for a given velocity field, in the magnetic diffusive regime.

\subsubsection{Torque steps, torsional oscillations}
\label{stepssection}

\begin{figure}
        \begin{center}
                \includegraphics[height=8.0 cm, keepaspectratio, angle=0]{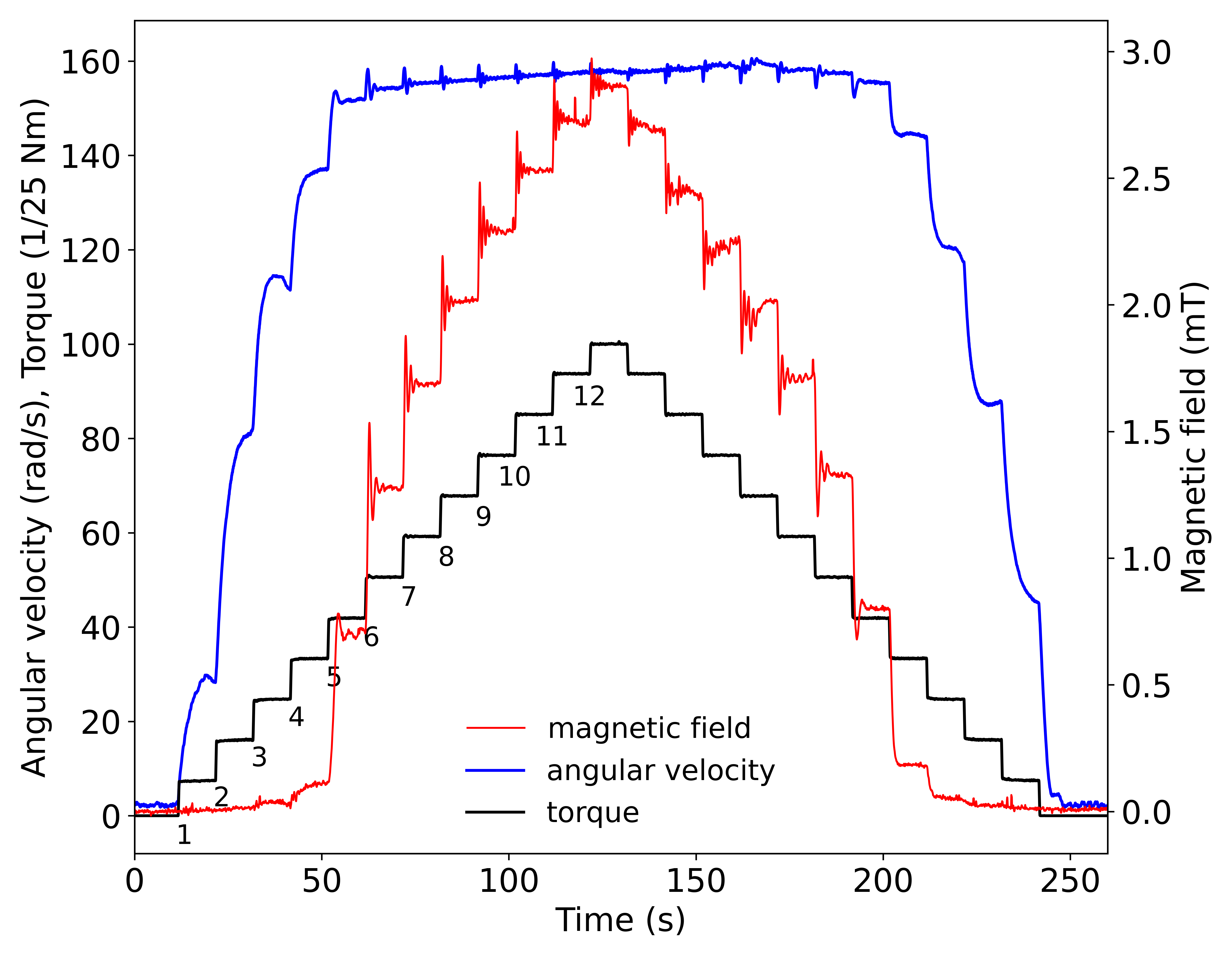}
                \caption{Angular velocity, torque, magnetic field in a single run with imposed steps of torque in the positive direction of rotation. }
                \label{stepsdynamo}
        \end{center}
\end{figure}

When the experimental setup Fury is run with an electric motor, we can impose a torque and step changes over a very short duration of less than $0.1$~s. In Fig.~\ref{stepsdynamo}, we show the time evolution of an experiment where we have imposed 12 steps of increasing torque (labeled 1 to 12) and the reverse 12 steps of deceasing torque, in black continuous line. Each step duration is 10~s and the maximum torque was 4~Nm. As a result of these torque changes, angular velocity (blue line) increases significantly for the 5 first steps, then remains globally at the same level and finally goes down in the 5 last steps. After step 5 and until just before step 20, self-exciting dynamo is at work, as shown by the value of the magnetic field (red curve), which maintains the angular velocity close to the dynamo threshold. The dynamo has actually perhaps started after the 4th step, as we can see a low level of magnetic field already. This is an indication that the dynamo threshold is not as sharp as expected for a kinematic dynamo, within a range of angular velocities, and we shall later explore this phenomenon.

\begin{figure}
        \begin{center}
                \includegraphics[height=8.0 cm, keepaspectratio, angle=0]{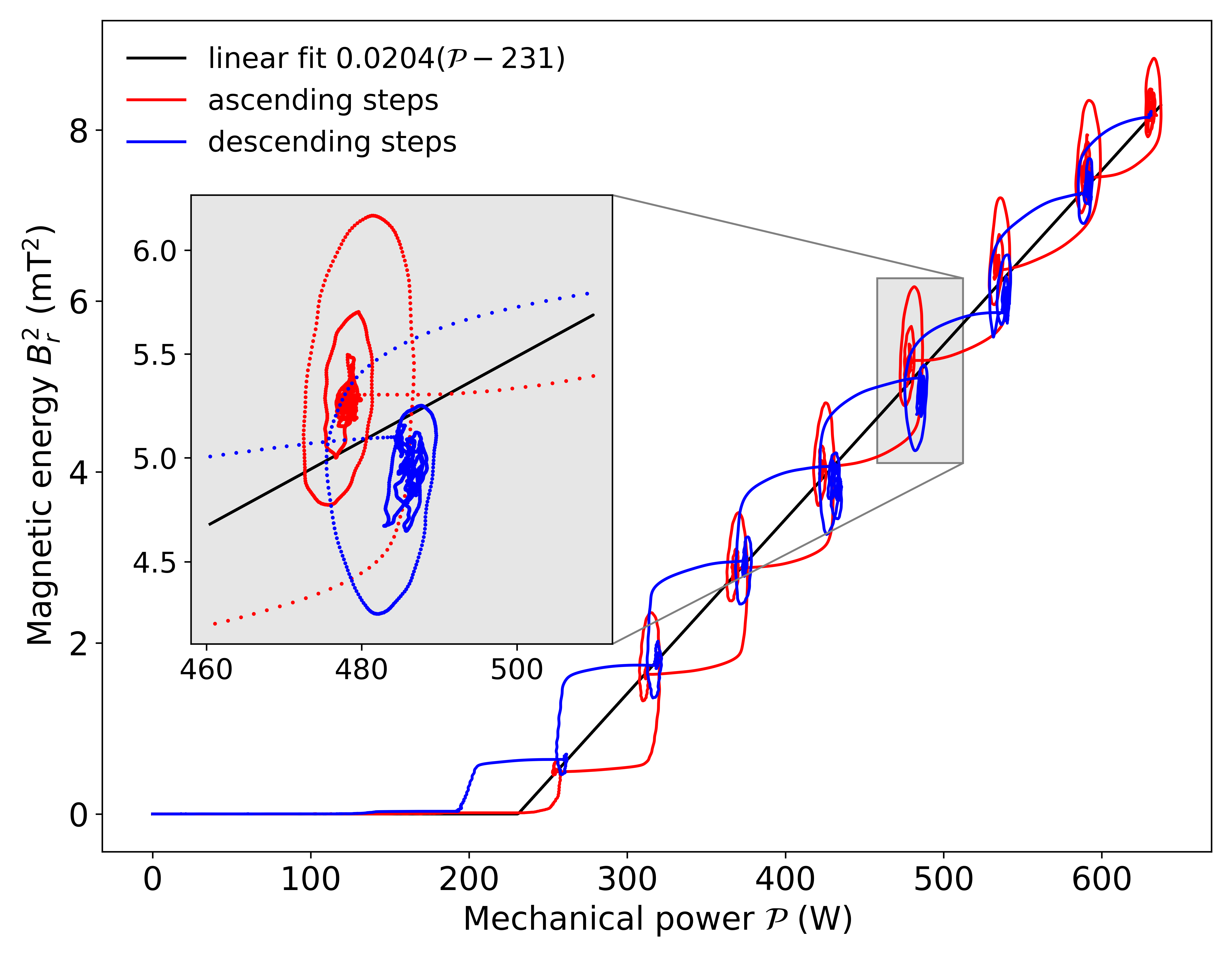}
                \caption{Square of the measured magnetic field $B_r^2$ as a function of the mechanical power input $\mathcal{P}$ during the run shown in Fig.~\ref{stepsdynamo}.}
                \label{stepsenergy}
        \end{center}
\end{figure}

In Fig.~\ref{stepsenergy}, the same experiment is represented in terms of mechanical energy input from the motor and square of the magnetic component $B_r$ on the external surface of the stator (position $z=3.7$~cm along the profile {$\tt{A}$-$\tt{B}$} in Fig.~\ref{mode}). In this diagram, the time evolution of the run is plotted in red while the torque is increased, then in blue for the second part of decreasing torque. In the dynamo regime, each step increase of the torque leads to an increase of mechanical power, followed by an increase of the magnetic energy. There is a transient phase of oscillations towards a final stationary regime where the path follows a converging spiral. The oscillations of the magnetic field can also be seen in Fig.~\ref{stepsdynamo} after each torque step. When the torque steps are in the decreasing phase (blue line), we observe a similar trend: the mechanical power goes down quickly and a phase of oscillating magnetic field follows with a final lower level of magnetic field. As mentioned above, the stationary points are well aligned along a straight line and we find a good agreement with the expression $B_r^2 = 0.0204 (\mathcal{P} -231 )$ above the dynamo threshold ($B_r$ in mT and the mechanical power input $\mathcal{P}$ in Watt).

\begin{figure}
        \begin{center}
                \includegraphics[height=8.0 cm, keepaspectratio, angle=0]{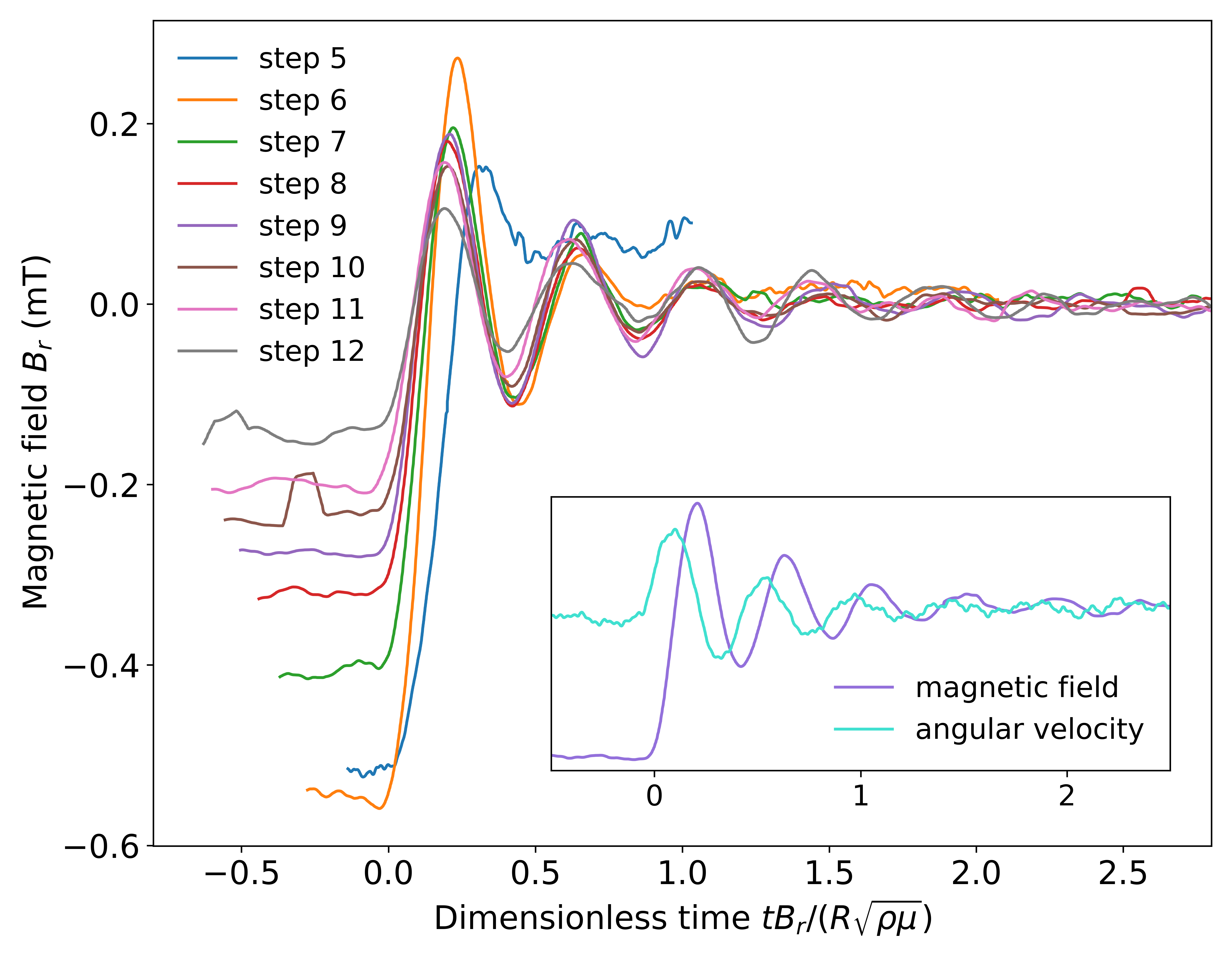}
		\caption{Stack of magnetic field oscillations after steps of torque (see Fig.~\ref{stepsdynamo} for the numbering of steps) with the dimensionless time scale $t B_r / (R \sqrt{\rho \mu })$. The inset (step 9) shows that the corresponding oscillations of angular velocity are in phase quadrature relative to magnetic oscillations. }
                \label{stepsoscill}
        \end{center}
\end{figure}

In Fig.~\ref{stepsdynamo}, it seems that the oscillations get faster as the magnetic field increases. We plot the oscillations following steps 5 to 12 in Fig.~\ref{stepsoscill}, using a dimensionless timescale proportional to the intensity of the magnetic field. Specifically, we use a timescale relevant to Alfv\'en waves, $t B_r / (R \sqrt{\rho \mu})$ with $R=0.05$~m the radius of the rotor, $\rho = 8960$~kg~m$^{-3}$ copper density, $\mu = 4\, \pi \, 10^{-7}$~H~m$^{-1}$ magnetic permeability and $B_r$ (in Tesla) the limit radial magnetic field after each step. The oscillations of $B_r$ superimpose nicely in this dimensionless timescale, and their period is of order one (close to $0.5$). In the inset of Fig.~\ref{stepsoscill}, we have plotted again the magnetic signal after one of the steps (step 9), along with the signal of angular velocity fluctuations. Those signals are in phase quadrature relative to each other, similarly to Alfv\'en waves \cite{Alboussiere2011}.
Because of the axisymmetry of the configuration and magnetic mode, these oscillations seem to be of the same nature of a particular kind of Alfv\'en waves, {\it i.e.} torsional oscillations. Such oscillations have been detected in the Earth's core \cite{Gillet2010} from geomagnetic data and also in an experiment with liquid sodium \cite{Tigrine2019}.

However, they are not Alfv\'en waves, nor associated torsional oscillations. Alfv\'en waves develop when kinematic and magnetic diffusion are negligible. Here, even with the highest magnetic field, the period of the oscillations is just under one second, while magnetic diffusion through 5~cm of copper is about 0.02~s. The oscillations are related to the dynamo mechanism itself: after a step, the system is pushed initially in the supercritical domain $Rm > Rm_c$ leading to a positive growth rate of the (global diffusive) magnetic field. The resulting increase of dissipative Lorentz forces brings the velocity down and the setup enters the subcritical domain $Rm < Rm_c$ with a deficit in Lorentz forces compared to the applied torque. The rotation increases and this is the start of a new period of the oscillations. The dynamic equation for the rotor rotation $\omega$ can be written approximately as follows
\begin{equation}
        \rho R^2 \frac{\md \omega}{\md t} \simeq - \sigma \omega R^2 B_r^2 - \mathcal{T}_v + \mathcal{T} ,\label{angular_momentum}
\end{equation}
where $- \sigma \omega R^2 B_r^2$ is the Lorentz torque, $-\mathcal{T}_v$ the viscous torque and $\mathcal{T}$ the constant torque provided by the motor, per unit volume. Assuming the viscous torque to be constant, taking the time derivative leads to
\begin{equation}
        \rho R^2 \frac{\md ^2 \omega}{\md t^2} \simeq - 2 \sigma \omega R^2 B_r \frac{\md B_r}{\md t} - \sigma R^2 B_r^2 \frac{\md \omega}{\md t}.\label{dangular_momentum}
\end{equation}
From our previous studies of the anisotropic dynamo, see equation (C17) of \cite{Alboussiere2020} and (4.18) of \cite{Plunian2022}, we know that the growth rate of the magnetic field is proportional to $\omega - \omega _0$ (with a constant of order unity) near the dynamo threshold $\omega _0$ for angular velocity. Substituting $\md B_r / \md t = (\omega - \omega _0 ) B_r$ in equation (\ref{dangular_momentum}) and dividing by $\rho R^2$, one obtains
\begin{equation}
        %\rho R \frac{\md ^2 \left(\omega - \omega _0 \right) }{\md t^2} = - \sigma R B^2_r \left[ 2 \omega _0 \left(\omega - \omega _0 \right) + \frac{\md \left(\omega - \omega _0 \right) }{\md t} \right].\label{oscillation_equation}
        \frac{\md ^2 \left(\omega - \omega _0 \right) }{\md t^2} \simeq - 2 \frac{\sigma B^2_r}{\rho } \omega _0 (\omega - \omega _0 ) - \frac{\sigma B^2_r}{\rho } \frac{\md (\omega - \omega _0 ) }{\md t} .\label{oscillation_equation}
\end{equation}
The last term on the right-hand side is negligible provided the timescale of the solution is longer than $\omega _0^{-1}$. Making this assumption leads to a simple harmonic equation, of associated timescale
 $\tau = \sqrt{\rho / \left( \sigma \omega _0 B_r^2 \right)}$ which is the geometric average of the (short) turnover timescale $\omega _0^{-1}$ and the (long) Joule damping time $\rho/ ( \sigma B_r^2 )$. It is longer than $\omega _0^{-1}$, hence our assumption is justified: it is also coherent with the smaller relative oscillations of angular velocity compared to those of the magnetic field in Fig.~\ref{stepsdynamo} after each step. This timescale $\tau$ can be expressed using $Rm_c = \mu \sigma \omega _0 R^2$ under the form
\begin{equation}
        \tau = \frac{1}{\sqrt{Rm_c}} \frac{R}{V_A}, \label{timescale}
\end{equation}
where $V_A = B_r / \sqrt{\rho \mu}$ is the Alfv\'en velocity. This timescale $\tau$ is the timescale we have used to plot Fig.~\ref{stepsoscill}, without the constant $1/\sqrt{Rm_c} \simeq 1/5$. Even though the physics is different from that of Alfv\'en waves, a similar timescale arises from the coupling between rotor inertia and dynamo growth rate.

\section{\label{bifurc} Nature of the bifurcation}

\begin{figure}
        \begin{center}
                \includegraphics[height=8.0 cm, keepaspectratio, angle=0]{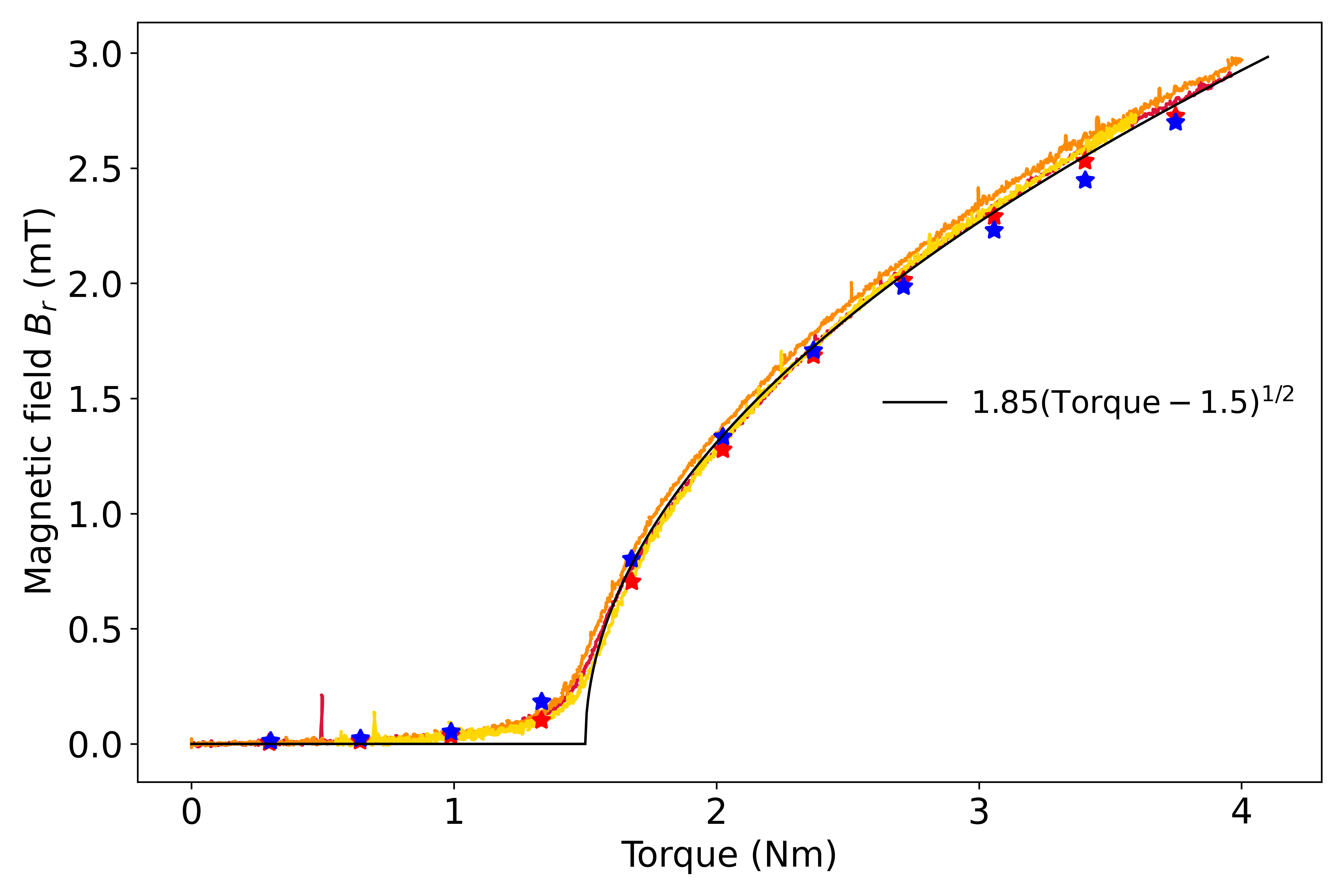}
		\caption{Measured magnetic field $B_r$ as a function of the torque input. The continuous coloured lines correspond to continuous variations of the torque shown in Fig.~\ref{couplevelocity} and the stars to the imposed torque steps shown in Fig.~\ref{stepsdynamo}, red (resp. blue) stars corresponding to increasing (resp. decreasing) torque steps.}
                \label{bif}
        \end{center}
\end{figure}

We plot in the same figure \ref{bif} different results concerning the dynamo bifurcation, as the measured magnetic field as a function of the torque imposed. This only concerns the 'dynamo' direction of rotation, since no magnetic field is generated in the other direction as expected from the theory. We replot the magnetic curves shown in Fig.~\ref{couplevelocity}, obtained with a (slow) continuous increase of the imposed torque. Then we also plot the steady value of the magnetic field reached after the transient period following each torque step shown in Fig.~\ref{stepsdynamo}. We observe that all measurements (continuous or steady) fall on a similar curve depicting an imperfect supercritical bifurcation, starting at a torque value of $1.5$~Nm with an exponent close to $1/2$ for the magnetic field. This bifurcation diagram is coherent with the evolution of magnetic energy as a function of mechanical power shown in Fig.~\ref{stepsenergy}, considering that the velocity remains constant near $150$~rad~s$^{-1}$ above the dynamo threshold. 

No clear hysteresis has been observed: increasing and decreasing steps of torque lead to points falling on the same curve. This applies also to the imperfect part of the bifurcation. Possible reasons for the imperfection could be related to the Earth ambiant magnetic field, or to small magnetic metallic pieces (bolts, ball bearings, gears) that have not been eliminated completely. Another observation is probably associated with the imperfect bifurcation: the magnetic field produced has always the same orientation whereas the opposite orientation should be as likely. The imperfect bifurcation is thought to drive the eigenmode in a definite orientation, not random.  

Concerning the critical exponent $1/2$ of the increase of the magnetic field above the critical torque threshold, it is directly related to the fate of the mechanical energy dissipated in the system, as discussed previously. This dynamo behaves very closely to a kinematic dynamo: the solid parts possess no freedom to deform and change the induced magnetic field, whereas the liquid part is restricted to a very narrow gap $0.5$~mm so that, even if changes occur they will have a very limited impact. This implies that the velocity does not change above the critical torque. Hence (turbulent) dissipation remains constant in the liquid gap\footnote{Turbulent friction (and dissipation) in the galinstan layer is not affected by the magnetic field unless $c_f Re^2/Ha^4 < 1$ according to \cite{al2000}. This would require the Hartmann number $Ha=\sqrt{\sigma/(\rho \nu )}Bh$ to be more than 100 times larger than it is estimated with a magnetic field of 3~mT.}. The extra mechanical power injected above threshold is entirely dissipated as Joule dissipation. As the form of the magnetic field remains the same, the magnetic field is proportional to the electric currents, which are proportional to the square root of Joule dissipation. With a nearly constant velocity, mechanical power is proportional to the torque, hence this explains why the magnetic field intensity grows like the square-root of the torque above threshold.

On the contrary, in Karlsruhe, Riga and Cadarache dynamos, the mean velocity of the flow can be increased above the threshold value for dynamo action. Also, the extra power above threshold goes mostly into hydrodynamical turbulent dissipation while a smaller part is dissipated through Joule effect. There have been arguments to suggest a scaling such as $B^2 \sim V-V_c$ \cite{Gailitis2003,Petrelis2001}, where $B$ is the magnetic field, $V$ the mean velocity and $V_c$ the dynamo threshold velocity. These arguments can be based on dimensional analysis or on the scaling of the Lorentz forces on the mean flow. Combined with a turbulent dissipative scaling $\mathcal{P} \sim V^3$ for the required mechanical power $\mathcal{P}$, the resulting scaling for the magnetic field is $B \sim \sqrt{\mathcal{P}^{1/3} -V_c}$ which can look like anything between $\mathcal{P}^{1/2}$ and $\mathcal{P}^{1/6}$ depending on the proximity to the threshold.    

Let us now compare the dynamo bifurcation observed in Fury to that observed in the Lowes and Wilkinson dynamo. Both dynamos require a similar power of the order of 200~W, but the dynamo of Lowes and Wilkinson had to be made in a magnetic material so that its dynamo threshold could be reached. In their first paper \cite{Lowes1963}, Lowes and Wilkinson show that they could increase the velocity of one of the cylinders above the critical velocity and that this led to an increase of the magnetic field intensity. This is different in Fury where the velocity cannot be made larger than the threshold velocity as the magnetic field would grow indefinitely in amplitude. In their second paper \cite{Lowes1968}, Lowes and Wilkinson observe oscillations of the magnetic field (and of the rotation rates of the cylinders) above dynamo threshold. Such oscillatory regimes have been found in more recent numerical simulations of the Herzenberg dynamo \cite{brandenburg1998}, but they could only be seen at a much larger magnetic Reynolds number. Lowes and Wilkinson admit that the magnetic material (saturation and remanence) might play a large role in the observed regimes. Their last sentence expresses their regrets that they had to use a magnetic material: "We hope by further improvement of the geometry to obtain a sufficiently small critical $R_m$ that will allows us to make a model from copper".

\section{\label{concl} Conclusions}

We have demonstrated, with the experiment Fury, that the concept of anisotropic conductivity could be used effectively to produce a self-generated dynamo. Without any magnetic materials, just copper, and a small quantity of galinstan for electrical contact, we have built a dynamo that can be human-powered. The characteristics of the setup correspond rather well to the finite-element calculations, in terms of dynamo threshold and eigenmode of the magnetic field. When abrupt changes in the driving torque are imposed, rotation and magnetic oscillations have been put in evidence, due to the linear dynamo growth rate dependence around the threshold angular velocity.

Our measurements indicate that the magnetic field is axisymmetric. This cannot be exactly true: at small scale (smaller than the distance between the grooves), our setup is just made of materials of isotropic heterogeneous conductivity. So Cowling's theorems apply and an axisymmetric field cannot be sustained by dynamo action. So azimuthal variations must exist but they are small and we have not measured them yet. On a meso-scale (larger than the distance between the groove and smaller than the setup) we model the configuration by a material of homogeneous anisotropic electrical conductivity. In that case, axisymmetric dynamo solutions exist. Our results prove that this modelling is adequate for Fury.

As we have seen, small scale conductivity heretogeneity can be modelled as electrical anisotropy in some cases, and this is true for our experiment Fury. However,  there are other cases of electrical heterogeneity that are definitely different. For instance a uniform flow next to a solid wall with spatial electrical conductivity modulation in the streamwise direction can lead to dynamo action \cite{busse92}. In this case, the wavelength of the modulation is important and cannot be made zero. In the limit of zero wavelength, this would indeed correspond exactly to an anisotropic electrical conductivity with the flow direction as a principal axis of anisotropy. Our model of sliding plates \cite{Alboussiere2020} shows that such a configuration does not produce a dynamo. In another case of heterogeneity \cite{petrelis2016} an equivalent $\alpha$-effect is produced but this requires velocity length-scales comparable to the length-scale over which electrical conductivity varies so that an anisotropic conductivity would not lead to the same results.    

In the recent years, a self-exciting dynamo with spiral arms was tested by Avalos-Z\'uñiga {\it et al.} \cite{raul}. In 2018 the dynamo threshold was reached and the results presented at the MHD days meeting in Dresden (16-28 Nov 2018) and more results have been provided to us in a private communication \cite{RaulPrivate}. This dynamo was designed according to the concept of Bullard's dynamo disc, however it seems that an anisotropic conductivity would be a faithful model for this setup, too \cite{priede2013}. Concerning future experiments, one can imagine a hybrid solid-liquid configuration, similar to Fury but with a larger gap of liquid sodium, or just a solid anisotropic rotor inside a pool of liquid sodium. The advantage of these configurations is that dynamo action would be guaranteed from the solid anisotropic part and that fluid dynamics would interact non-linearly with the magnetic field. Considering the difficulties encountered to produce and study interesting non-linear dynamos, this alternative might be valuable.

The name Fury was given to the setup because this is a dynamo that should have the property of a dynamo growth-rate increasing without bound above the dynamo threshold, which we called a ``very fast'' or ``furious'' regime \cite{Alboussiere2020,Plunian2022}. That property is rather difficult to measure in the experiment because it would require a lot of mechanical power (within the resistance limits of the setup) to bring the rotor at a velocity well above the threshold velocity.  \\

{\bf Acknowledgements:} {We wish to acknowledge the participation of students of geology in Lyon, Lucie Petit, Dana Romanoz and Paul Gomez in the measurement of the magnetic profile of the dynamo mode. Thanks are due to Andy Jackson for suggesting the name 'Fury' for the setup.} \\
	{We thank the Programme National de Plan\'etologie (PNP) of CNRS/INSU, co-funded by CNES. The authors are grateful to the LABEX Lyon Institute of Origins (ANR-10-LABX-0066) Lyon for its financial support within the Plan France 2030 of the French government operated by the National Research Agency (ANR).}\\

{\bf Statements:}
{TA and MM designed the setup. TA and FP caried out the experiments. TA performed the data analysis and drafted the manuscript. All authors read and approved the manuscript.}\\
{The authors declare that they have no competing interests.}

\bibliography{refs}% Produces the bibliography via BibTeX.

\end{document}